\documentclass[12pt, letterpaper]{article}
\usepackage[top=1in, bottom=1.5in, left=1in, right=1in]{geometry}

\usepackage[utf8]{inputenc}
\usepackage{booktabs}
\usepackage{hyperref}
\usepackage{aas_macros}
\usepackage{color}
\usepackage{graphicx}
\usepackage{caption}
\usepackage{lineno}
\usepackage{deluxetable}

\title{Mini-survey of the northern sky to Dec $<+30$}
\author{P. Capak, 
D. Sconlic, 
J-C. Cuillandre,
F. Castander,
A. Bolton,\\
R. Bowler,
C. Chang,
A. Dey,
T. Eifler,
D. Eisenstein,
C. Grillmair, \\
P. Gris,
N. Hernitschek,
I. Hook,
C. Hirata,
B. Jain
K. Kuijken,\\
M. Lochner,
J. Newman,
P. Oesch,
K. Olsen,
J. Rhodes, \\
B. Robertson,
D. Rubin,
C. Scarlata, 
J. Silverman,
S. Wachter,\\
Y. Wang,
The Tri-Agency Working Group}
\date{November 2018}

\begin{document}

\maketitle

\begin{abstract}
We propose an extension of the LSST survey to cover the northern sky to DEC $< +30$ (accessible at airmass $<1.8$).  This survey will increase the LSST sky coverage by $\sim9,600$ square degrees from 18,900 to 28,500 square degrees (a 50\% increase) but use only $0.6-2.5\%$ of the time depending on the synergies with other surveys.  This increased area addresses a wide range of science cases that enhance all of the primary LSST science goals by significant amounts.  The science enabled includes: increasing the area of the sky accessible for follow-up of multi-messenger transients including gravitational waves, mapping the milky way halo and halo dwarfs including discovery of RR Lyrae stars in the outer galactic halo, discovery of $z>7$ quasars in combination with \emph{\emph{Euclid}}, enabling a second generation DESI and other spectroscopic surveys, and enhancing all areas of science by improving synergies with \emph{Euclid}, WFIRST, and unique northern survey facilities.  

This white paper is the result of the Tri-Agency Working Group (TAG) appointed to develop synergies between missions and presents a unified plan for northern coverage.  The range of time estimates reflects synergies with other surveys.  If the modified DESC WFD survey, the ecliptic plane mini survey, and the north galactic spur mini survey are executed this plan would only need 0.6\% of the LSST time, however if none of these are included the overall request is 2.5\% of the 10 year survey life. In other words, the majority of these observations are already suggested as part of these other surveys and the intent of this white paper is to propose a unified baseline plan to carry out a broad range of objectives to facilitate a combination of multiple science objectives.  A companion white paper gives \emph{Euclid} specific science goals, and we support the white papers for southern extensions of the LSST survey. We also endorse the white papers from the LSST Dark Energy Science Collaboration (DESC) arguing for modifications to the Wide Fast Deep (WFD), and the Big Sky white paper.

\end{abstract}

\clearpage

\section{White Paper Information}
Peter Capak, capak@caltech.edu
Dan Scolnic, dscolnic@kicp.uchicago.edu

\begin{enumerate} 
\item {\bf Science Category:} All science themes are improved by this survey.  This survey also improves synergies with \emph{Euclid}, WFIRST, and all northern facilities.
\item {\bf Survey Type Category:} Mini Survey 
\item {\bf Observing Strategy Category:} 
      An integrated program with science that hinges on the combination of pointing and detailed 
	observing strategy  - we propose a wide range of science than can be added or enhanced by a modest extension to the northern sky.  We propose a loose cadence in some bands to ensure proper motions and RR Lyrae periods are well sampled. 
\end{enumerate}

\clearpage

\section{Scientific Motivation}

Adding the accessible northern sky to the LSST survey area enhances the science for three distinct reasons: 1) The increase in sky area improves the statistical power of sky limited surveys such as transient, cosmology, and milky way halo surveys; 2) The northern sky contains unique objects, such as the Virgo and Coma clusters as well as the Andromeda group;   and 3) It optimizes synergistic science with space-based and northern facilities that can not observe the best LSST sky. Here we enumerate a range of science goals well aligned with the LSST Science categories that are enhanced with a northern extension to the LSST survey.\\

{\noindent \bf Multi-Messenger Transient Follow-up:}
The discovery of gravitational waves by LIGO has opened up a new observational window for astrophysics \cite{Abbot17}.  We are now in an era of multi-messenger astrophysics where transient phenomenon can be discovered not only from electromagnetic radiation but from neutrinos, cosmic rays, and gravitational waves. However, these other sources of information have poor localization, are uniformly distributed on the sky, and provide limited information on the nature of the source in question.  As a result, rapid follow-up of a large area of the sky, most efficiently done with LSST, is needed to find optical counterparts. Obtaining a template of the northern sky, needed for image subtraction and candidate identification, will increase the area for transient searches with LSST by $50\%$ with a $1.5-2.3$~ABmag template sensitivity increase over Pan-STARRS. The template depth will likely be the limiting factor of the search efficiency.\\

{\noindent \bf The Local Group:}
Substructure in the Milky Way and our neighboring galaxies contains significant information on how our galaxy formed and the nature and distribution of dark matter.  Covering the northern sky increases the volume probed by LSST and includes a key piece of the structure: the Andromeda (M31) group.  A relatively shallow survey with epochs spread over 10 years and a cadence in $g$ band designed to characterize RR Lyrae stars would provide significant science returns.  When paired with a survey of the Magellanic clouds it would provide a consistent distance ladder out to our nearest galactic neighbors.   The most valuable part of this survey is the overlap with \emph{Euclid}, which will enable more than an order of magnitude increase in the surface brightness sensitivity to streams and dwarfs by improving star/galaxy separation.  Furthermore, the northern sky is accessible to the Dark Energy Spectroscopic Instrument (DESI) and Subaru Prime Focus Spectrograph (PFS) projects which could carry out spectroscopic follow-up of stars to create dynamical maps of the local universe. \\

{\noindent \bf The nearby ($z<0.1$) Universe:}
Several of the largest structures in the nearby universe, including the Coma and Virgo clusters, are in the northern sky.  With a modest extension to the north, LSST would map these structures and $\sim50\%$ more of the local volume, enabling a wide range of environment studies of galaxy formation.  Specifically, LSST imaging which would resolve the morphology, star formation, and stellar mass down to low surface brightness limits and find globular clusters and other substructure around galaxies as a function of local environment. Furthermore, Coma and part of Virgo overlap with the \emph{Euclid} Survey which will provide $0.1^{\prime\prime}$ spacial resolution 0.6-2$\mu$m imaging and $1.1-1.8\mu$m $R\sim250$  spectroscopy of these environments. These data will provide detailed structural and dynamical measurements that can be combined with the LSST mass and star formation estimates. \\

{\noindent \bf Cosmology:}
A detailed science case for LSST-\emph{Euclid} cosmology is given in \cite{LSST-Euclid}.  In brief, the main gain to cosmology science comes from a combination of better photometric redshifts, improved weak lensing shear measurements, and improved de-blending of galaxy photometry.  Figure \ref{fig:pz-improve} shows an estimate of the improvement in photometric redshifts based on real data from the CFHT and VISTA telescopes and representative spectra from the C3R2 survey \cite{c3r2}. A further enhancement will come from observing a larger area that overlaps with the Dark Energy Spectroscopic Instrument (DESI) survey and can be observed by the Subaru Prime-Focus Spectrograph; both  are  in the north.  The DESI data can be used to calibrate photometric redshifts for weak lensing measurements and conduct joint cosmology probes, while both DESI and PFS can carry out dedicated follow-up observations.  Finally, the \emph{Euclid} area of the sky is chosen to be optimal for weak lensing from space.  If sufficiently deep LSST data were obtained over this area, WFIRST could conduct a follow-on survey of this sky area in the broad W band to further enhance cosmological measurements by increasing the sky coverage. \\

{\noindent \bf The High-Redshift ($z>1.5$) Universe:}
By extending the LSST survey to the northern areas as also covered by \emph{Euclid} LSST will extend the region with near-infrared-based stellar mass estimates from $z\sim1.3$ to $z\sim3.5$ over $40\%$ more volume that covered by the current WFD survey.  This will enable studies of galaxy evolution beyond the peak of the global star formation rate at $z\sim2-3$.  Furthermore, the addition of \emph{Euclid} spectra will provide near-infrared spectral line measurements for many LSST objects.  The combination of \emph{Euclid} and LSST data will also enable the discovery of AGN and lensed/luminous galaxies at $7<z<12$ to enable detailed studies of the reionization epoch. Based on current estimates \cite{bouwens,oesch,ono} we expect between $30-500$ galaxies and $\sim70$ AGN per 1000 square degrees.  Neither LSST or \emph{Euclid} can do this alone, so only by combining data will these studies be possible.\\

{\noindent \bf Follow-up Facilities in the North:}
Major and unique current and up-and-coming multi-wavelength survey facilities and instruments will be available in the northern hemisphere that are unlikely to be duplicated in the south before the LSST survey is completed.  Maximizing overlap with these facilities will naturally lead to enhanced LSST science. For instance, the DESI and Subaru PFS spectroscopic facilities should provide the most powerful wide-field spectroscopic capabilities in the next decade.  Subaru Hyper Suprime-Cam (HSC) and CFHT can conduct surveys in intermediate- and narrow-band filters that enable fundamentally new types of survey science.  In the sub-millimetre, the 50\,m LMT, the 30\,m IRAM, and 15\,m JCMT telescopes provide the only high-sensitivity/resolution sub-mm survey capability for the next decade with no comparable facility available in the southern hemisphere. Finally, Keck, Gemini North, and the Thirty Meter Telescope are all located in the north.

\begin{minipage}{0.99\columnwidth}
\includegraphics[width=1\columnwidth]{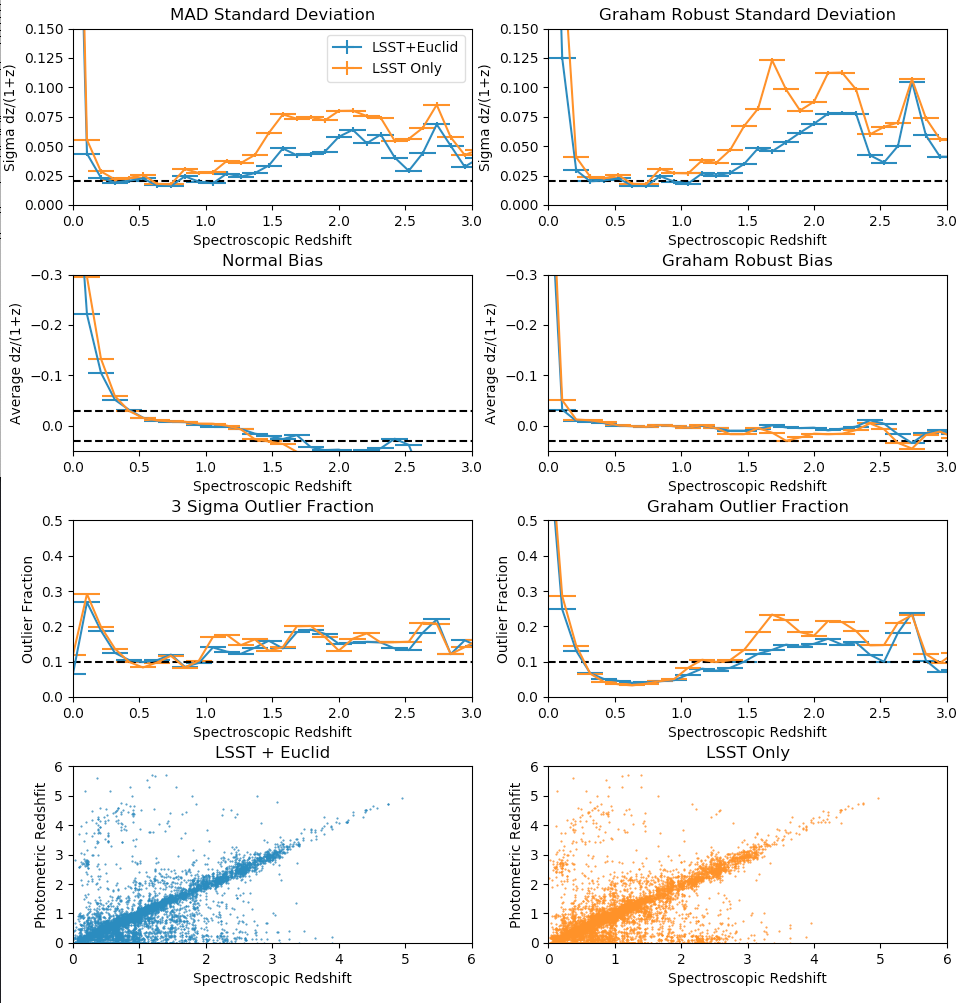}
\captionof{figure}{A comparison of spectroscopic and photometric redshifts for a representative set of 25,092 high-quality spectroscopic redshifts from the C3R2 survey \cite{c3r2} with real data comparable to that which will be obtained with LSST in the Wide-Fast-Deep (WFD) and \emph{Euclid} wide surveys.  It is clear that combining \emph{Euclid} and LSST data significantly improves the photometric redshift performance.  We show standard photometric redshift performance metrics as well as those used by \cite{graham2017} for LSST.
The optical imaging comes from CFHT-LS deep fields in u,g,r,i,z that are deeper than those that will be obtained by the LSST WFD survey.  The CFHT-LS photometry were degraded to the expected WFD depth in consultation with the LSST project.  The near-infrared imaging comes from the VISTA VIDEO and UltraVISTA surveys in Y,J,H,K bands that are comparable to what will be obtained by \emph{Euclid} in the near-infrared. }  
\label{fig:pz-improve}
\end{minipage}

\pagebreak
\section{Technical Description}

\subsection{High-level description}
To achieve all of the science we propose to observe the $\sim$9,600 square degrees of sky at $+2<$DEC$<+30$ with several epochs spread over the 10 year LSST survey. This would add to the $\sim18,900$ square degrees in the existing LSST baseline for a $50\%$ increase in area.  For planning and metric purposes we split this area into three sections listed in Table \ref{tab:SurveyFootprint} with each area having a different coverage and cadence designed to meet the overall science goals.  These sections are each synergistic with other proposed mini-survey which we are aware of and the split was in part designed to simply combine the suggested mini-surveys. In parallel to the three survey sections we enumerate the requirements of each science goal and which sections of the survey they use starting with Table \ref{tab:footprint}.  This is intended to facilitate understanding whether science will be lost by deviating from our proposed depth/cadence.\\

In the following list we give an overview of the requirements of each science case:

\begin{itemize}
\item Imaging the northern sky will increase the available area for multi-messenger follow-up by 50\%.  Based on the LSST science book, follow up of multi-messenger transients requires a template capable of characterizing the 24th magnitude sources expected to be the counterparts of a-LIGO gravitational wave events \cite{ScolnicK18}.  Since this is within the single-visit depth, the main requirement is at least three visits per band in all bands to fill in chip gaps and be robust to artifacts such as cosmic rays and scattered light in the images. This robustness is needed to provide a well-calibrated template for TOO observations.  The existing Pan-STARRS data are $1.5-2.3$~ABmag shallower than even single visit LSST depths and are thus not suitable for use as templates, and additionally have different filters which would cause issues in the image subtraction pipeline.  These data should be obtained early in LSST observing.

\item Following the LSST science book, measurements of galactic streams requires coverage in at least $u$, $g$, and $i$ bands to characterize the metallicities and classes of stars and ideally also r and z bands to improve classification. At least three epochs, each with two visits  in $g$ and $r$ bands should be spread over the 10 years of the LSST survey to measure proper motions and create a uniform mosaic.  In the areas where LSST and \emph{Euclid} overlap, the integrated depth in $u$, $g$ and $i$ should be matched to the \emph{Euclid} depth of $r,i<25.2$ which will improve star/galaxy separation, allowing for more than an order of magnitude improvement in surface brightness sensitivity in the areas where \emph{Euclid} data are available.  

\item Following the work in Pan-Starrs \cite{RRLyrae}, identifying and characterizing RR Lyrae stars in the outer halo of the Milky Way and in galactic streams requires $\sim10$ single visit observations in at least $g$ and $r$ bands randomly spread over the 10 year lifetime. 

\item Studies of resolved galaxies in the nearby universe should match the \emph{Euclid} depths of $RIZ<25.2$ to maximize the science return from combined \emph{Euclid} and LSST data.

\item Joint LSST/DESI observations will enhance cosmological measurements and studies of dark matter with joint redshift-space distortion (RSD) and weak lensing measurements. The Ultraviolet Near-Infrared Optical Northern Survey (UNIONS, by CFHT and Pan-STARRS), which is imaging the 5,000 square degree northernmost \emph{Euclid} area at DEC$>+30$ estimates this will require depths of $u<23.6$, $g<25.7$, $r<25.1$, $i<24.8$, $z<24.6$.  This survey also requires WFD like image quality in $r$ and $i$ bands to measure weak lensing shapes.  Our proposed northern extension would increase the area available for this type of analysis by a factor of $\sim2$.

\item The DESI survey is expected to be completed in 2025, after which the facility would likely be available for a second-generation survey.  The science plan for such a survey has not yet been decided, but several options are being studied.  An example is to use Lyman Break Galaxies (LBGs) to study large-scale structure at $z>2$.  To do so with a surface density of $\sim2000$ per square degree requires selecting objects to $\sim24$~ABmag \cite{bouwens,oesch,ono}.   To ensure uniform selection they require a $10\sigma$ detection in the selection bands at $\sim24$~ABmag and $2\sigma$ detection at 2 magnitudes fainter in the dropout bands to ensure a robust LBG/photo-z selection \cite{bouwens,oesch,ono}.  This translates to an overall requirement of $5\sigma$ detections in $u,g,r,i<25.0$, and $z<24.6$.  Alternative strategies focusing on Lyman-$\alpha$ emitters would require even deeper selection imaging, as the equivalent widths increase in fainter galaxies. 

\item  To optimize synergies with \emph{Euclid} a depth of $g<25.7, r<25.1, i<24.8, z<24.6$ should be obtained over the areas that overlap with \emph{Euclid}.  Adding $u<25.4$ and going up to 0.8 magnitudes deeper than these limits will continue to improve cosmological constraints.  Reaching these larger depths would also enable a possible follow-on WFIRST wide-area weak lensing survey that could further increase cosmological measurements.   The details of these improvements are outlined in a companion white paper.  
\end{itemize}

\subsection{Footprint -- pointings, regions and/or constraints}

\includegraphics[width=1.0\columnwidth]{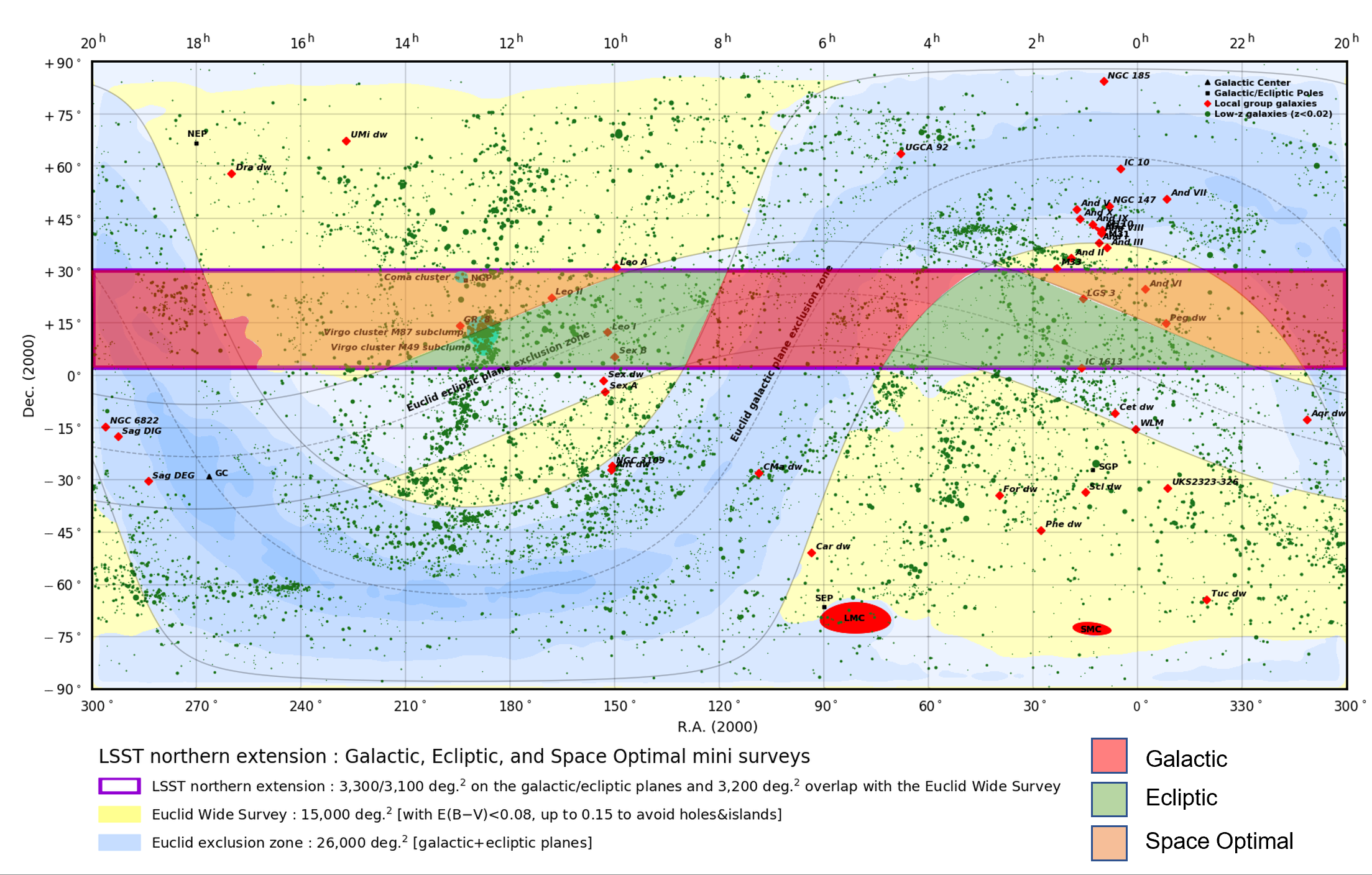}
\captionof{figure}{The proposed survey footprint is shown as a purple box with a combined stellar density and extinction map from the Milky Way shaded in the background.  We break this area into three areas with different cadences and depths listed in Tables \ref{tab:SurveyFootprint}, \ref{tab:NvisitSurvey} and \ref{tab:timeline}.  Known Local Group galaxies are marked in red and known galaxies at $z<0.02$ are marked in green with the Virgo and Coma clusters marked in cyan.  Areas of the sky that will be covered by the \emph{Euclid} survey are highlighted in yellow.  \label{Fig:footprint}
}
\begin{deluxetable}{ccccc}
\tabletypesize{\footnotesize}
\tablewidth{0pt}

 \tablecaption{ Survey Footprints \label{tab:SurveyFootprint}}
 \tablehead{
 \colhead{Survey} & \colhead{Area (Deg$^2$)} & \colhead{N} & \colhead{Notes} & \colhead{Synergistic}\\
  \colhead{} & \colhead{} & \colhead{Visits} & \colhead{} & \colhead{Mini-Surveys}
 }
 \startdata 
Galactic & 3,300  & 32 & Area at galactic latitude $b<25$ & a,e\\
Ecliptic  & 3,100 & 46 & Low Extinction area at ecliptic latitude $<15$ & b,c,e\\
Space Optimal & 3,200  & 64 & \emph{Euclid} area of the sky at at $+2<$DEC$<+30$ & b,d,e \\
\enddata
{ {\bf a)} North Galactic Spur. {\bf b)} DESC WFD proposal. {\bf c)} Ecliptic plane survey. {\bf d)} \emph{Euclid} Synergy. e) Big Sky.}
\end{deluxetable}

The areas of the sky at $+2<$DEC$<+30$ we are considering for our survey are listed in Table \ref{tab:SurveyFootprint} and illustrated in Figure \ref{Fig:footprint}. We have broken this area into three regions, the galactic plane region, the extragalactic regions near the ecliptic that are not optimal for space-based observations, and the Space-Optimal regions observed by \emph{Euclid}.  The areas relevant for each science case are listed in Table \ref{tab:footprint}.\\

We note the proposed survey overlaps significantly with many other proposed mini-surveys.  With the exceptions of the specific cadences listed, the imaging obtained for those surveys will be sufficient for this program.  The synergistic mini-surveys are noted in Table \ref{tab:SurveyFootprint}.

\begin{deluxetable}{ccc}
\tabletypesize{\footnotesize}
\tablewidth{0pt}

 \tablecaption{ Science Cases Linked to Footprints \label{tab:footprint}}
 \tablehead{
 \colhead{Science Goal} & \colhead{Area (Deg$^2)$} & \colhead{Surveys used}
 }
 \startdata 
TOO Template & 9,600 &  All \\
Local Group & 9,600 &  All\\
Galactic RR Lyrae & 9,600 & All\\
Nearby Universe & 6,300  & Ecliptic, Space Optimal\\
\emph{Euclid}/WFIRST synergy & 3,200 & Space Optimal\\
DESI Synergy & 6,300  & Ecliptic, Space-Optimal\\
DESI-2 &  6,300  & Ecliptic, Space-Optimal\\
High-z &  3,200 & Space Optimal\\
\enddata
\end{deluxetable}

\subsection{Image quality}

For optimal synergy we recommend following the image quality requirements of the Wide-Fast-Deep survey.  The most demanding requirement comes from the LSST-DESI weak lensing synergy which places strict requirements on $r$ and i band image quality. To produce the highest quality images, we recommend scanning along the meridian as much as possible. \\

We recognize that atmospheric dispersion will become problematic at the highest airmasses. This may result in degraded sensitivity and require more visits than those requested here. 

\subsection{Individual image depth and/or sky brightness}
None.

\pagebreak

\subsection{Co-added image depth and/or total number of visits}
The depth/visit requirements for each science case are listed in Table \ref{tab:Nvisit} and combined into a unified survey plan in Table \ref{tab:NvisitSurvey}.  The Table \ref{tab:Nvisit} numbers are thresholds for each science case and are not cumulative.\\

Some specific notes for each case are:

\begin{itemize}
\item The TOO Template is driven by the number of visits required to provide a robust template.  
\item The Local Group science case is driven by the number of visits required for proper motion measurements.  
\item The galactic stream science case requires a depth of $u,g,i > 25.2$ at $5\sigma$ to match the \emph{Euclid} point source sensitivity. 
\item The RR Lyrae cases are driven by the need for 10 epochs needed to measure periods.
\item The \emph{Euclid} science case require a minimum of five exposures to produce uniform, systematics-controlled photometry and are otherwise driven by sensitivity requirements, not by numbers of visits.
\item The high-z case is driven by the sensitivity in z band needed to select quasars at $z>7$.
\end{itemize}

\begin{deluxetable}{ccccccccccccc}
\tabletypesize{\footnotesize}
\tablewidth{0pt}

 \tablecaption{ Depth/Number of Visits Required for Each Science Case \label{tab:Nvisit}}
 \tablehead{
 \colhead{Science Goal} & \colhead{u} & \colhead{g} & \colhead{r} & \colhead{i} & \colhead{z} & \colhead{y} & \colhead{u} & \colhead{g} & \colhead{r} & \colhead{i} & \colhead{z} & \colhead{y}
 }
 \startdata 
TOO Template & N/A & N/A & N/A & N/A & N/A & N/A & 3 & 3 & 3 & 3 & 3 & 3  \\
Local Group & N/A & N/A & N/A & N/A & N/A & N/A & 3 & 6 & 6 & 3 & 3 & 0  \\
Galactic RR Lyrae & N/A & N/A & N/A& N/A & N/A & N/A & 0 & 10 & 10 & 0 & 0 & 0  \\
Nearby Universe & 25.0 & 25.4 & 25.7 & 24.8 & 24.6 & N/A & 8 & 3 & 3 & 5 & 10 & 0  \\
\emph{Euclid} Synergy & 25.4\tablenotemark{*} & 25.7 & 25.1 & 24.8 & 24.6 & N/A & 15\tablenotemark{*} & 5 & 5 & 5 & 10 & 0  \\
DESI Synergy & N/A & 25.7 & 25.1 & 24.8 & 24.6 & N/A & 0 & 3 & 3 & 5 & 10 & 0 \\
DESI-2 & 25.0 & 25.0 & 25.0 & 25.0 & 24.6 & N/A & 8 & 3 & 3 & 5 & 10 & 0  \\
High-z Quasars & N/A & N/A & N/A & N/A & 25.0 & N/A & 0 & 0 & 0 & 0 & 21 & 0  \\
\enddata

\tablenotetext{*}{Optimal but not required.}
\end{deluxetable}

\begin{deluxetable}{cccccccccccccc}
\tabletypesize{\footnotesize}
\tablewidth{0pt}

 \tablecaption{ Science Compilation : Depth/Number of Visits per Survey \label{tab:NvisitSurvey}}
 \tablehead{
 \colhead{Survey} & \colhead{Area} & \colhead{u} & \colhead{g} & \colhead{r} & \colhead{i} & \colhead{z} & \colhead{y} & \colhead{u} & \colhead{g} & \colhead{r} & \colhead{i} & \colhead{z} & \colhead{y}
 }
 \startdata 
Galactic & 3,300 & 24.5 & 26.1 & 25.7 & 24.6 & 23.9 & 22.7 & 3 & 10 & 10 & 3 & 3 & 3 \\
Ecliptic & 3,100 & 25.0 & 26.1 & 25.7 & 25.4 & 24.8 & 22.7 & 8 & 10 & 10 & 5 & 10 & 3 \\
Space Optimal & 3,200 & 25.4\tablenotemark{*} & 26.1 & 25.7 & 25.4 & 25.3 & 22.7 & 15\tablenotemark{*} & 10 & 10 & 5 & 21 & 3 \\
\enddata
\tablenotetext{*}{Optimal but only $u<25.0$ $\sim8$ visits required.}
\end{deluxetable}

\subsection{Number of visits within a night}
No requirements.

\subsection{Distribution of visits over time}
The distribution of visits over time for each science case is listed in Table \ref{tab:Dvisit}.  The main requirements are:

\begin{itemize}
\item The TOO Template should be obtained early. 
\item The Local Group science case requires 3 visits with the maximum time baseline possible to measure proper motions.  Ideally these would be in year 1, 5, and 10.  
\item The RR Lyrae cases are driven by the need to sample the light curves as well as possible at 10 epochs and rely on the buildup of small phase errors over the length of the survey.
\item \emph{Euclid} requires imaging when it observes that area of the sky which will be between 2025 and 2028.
\item DESI requires the imaging before the start of DESI-2 which would be by the end of 2024.
\end{itemize}

\begin{deluxetable}{lccl}
\tabletypesize{\footnotesize}
\tablewidth{0pt}

 \tablecaption{ Distribution of Visits \label{tab:Dvisit}}
 \tablehead{
 \colhead{Survey} & \colhead{Number of} & \colhead{Time Between } & \colhead{Required}\\
  \colhead{Goal} & \colhead{Epochs} & \colhead{Epochs} & \colhead{By}
 }
 \startdata 
TOO Template & 1 & N/A & Start of survey \\
Local Group & 3 & $>2$ years & End of survey\\
Galactic RR Lyrae & Spread over 10 years &  & End of survey\\
Nearby Universe & N/A & N/A & N/A \\
\emph{Euclid} Synergy & N/A & N/A & 2025-2028 \\
DESI & N/A & N/A & 2024 \\
\enddata
\end{deluxetable}

\begin{deluxetable}{ccccccccccccc}
\tabletypesize{\footnotesize}
\tablewidth{0pt}

 \tablecaption{ Proposed Survey Time Line \label{tab:timeline}}
 \tablehead{
 \colhead{LSST} & \colhead{Year} & \colhead{u} & \colhead{g} & \colhead{r} & \colhead{i} & \colhead{z} & \colhead{y} & \colhead{Total} & \colhead{Notes} & \colhead{Percent\tablenotemark{h}}\\
 \colhead{Year} & \colhead{} & \colhead{} & \colhead{} & \colhead{} & \colhead{} & \colhead{} & \colhead{} & \colhead{} & \colhead{} & \colhead{of Year}
 }
 \startdata 
1     & 2023 & 3,3,3 & 3,3,3 & 3,3,3 & 3,3,3 & 3,3,3 & 3,3,3 & 18,18,18 & a,d   & 9.3/2.0\\
2     & 2024 & 0,2,2 & 0     & 0     & 0,1,1 & 0,4,4 & 0     & 0,7,7    &       & 2.4/0.8\\
3     & 2025 & 0,3,3 & 0     & 0     & 0,1,1 & 0,3,3 & 0     & 0,7,7    & b     & 3.1/0.8 \\
4     & 2026 & 0,0,3 & 1,1,1 & 1,1,1 & 0     & 0,0,3 & 0     & 2,2,8    & c,d   & 2.1/0.9\\
5     & 2027 & 0,0,0 & 2,2,2 & 2,2,2 & 0     & 0,0,0 & 0     & 4,4,4    & c,e   & 2.1/0.5\\
6     & 2028 & 0,0,4 & 0     & 1,1,1 & 0     & 0,0,4 & 0     & 1,1,8    & d,e,f & 1.7/0.9\\
7     & 2029 & 0     & 1,1,1 & 0     & 0     & 0,0,4 & 0     & 1,1,4    & d,e   & 1.0/0.5\\
8     & 2030 & 0     & 1,1,1 & 0     & 0     & 0     & 0     & 1,1,1    & d     & 0.5/0.1\\
9     & 2031 & 0     & 0     & 1,1,1 & 0     & 0     & 0     & 1,1,1    & d     & 0.5/0.1\\
10    & 2032 & 0     & 2,2,2 & 2,2,2 & 0     & 0     & 0     & 4,4,4    & d,g   & 2.1/0.5\\
Total &      & 3,8,15&10,10,10&10,10,10&3,5,5& 3,10,21&3,3,3 & 32,46,64 & & 2.5/0.7\\
\enddata
\tablenotetext{a}{TOO template required in first year, could also be done as part of science verification.}
\tablenotetext{b}{Start of DESI-2, if LSST starts earlier or if DESI is late observations before this date could be spread out.}
\tablenotetext{c}{High-z Z band imaging added, could be spread out more if required.}
\tablenotetext{d}{RRLyrae observations in g and r.}
\tablenotetext{e}{Second epoch of proper motions.  Two exposures to match \emph{Euclid} sensitivity of $r<25.2$.}
\tablenotetext{f}{\emph{Euclid} needs imaging.}

\tablenotetext{g}{Final epoch of proper motions.}
\tablenotetext{h}{The first number assumes only the baseline WFD survey to DEC$<+2$.  The second number assumes the DESC proposed WFD survey, the galactic plane mini-survey, and the ecliptic plane survey.}
\end{deluxetable}

\subsection{Filter choice}
These are listed in the other sections above. 

\subsection{Exposure constraints}
None.

\subsection{Other constraints}
To optimally facilitate \emph{Euclid} the \emph{Euclid} portions of the survey should be completed by 2028.  To faciliate DESI, the DESI depth should be reached by 2024.

\subsection{Estimated time requirement}
We estimated the time needed for each survey based on the DESC WFD white paper simulations, scaling to the number of total visits and area of each survey.  We assume the WFD survey has 825 visits per pointing, covers 18,000 square degrees, and uses 80\% of the total survey.  Table \ref{tab:time} gives the estimated survey time for each of the proposed surveys and the total at the bottom.

\begin{deluxetable}{ccccccc}
\tabletypesize{\footnotesize}
\tablewidth{0pt}

 \tablecaption{Time Estimates for Each Survey\label{tab:time}}
 \tablehead{
 \colhead{Survey} & \colhead{Total Visits} & \colhead{Area} &  \colhead{Fractional} & \colhead{Synergistic Mini}\\ 
  \colhead{} & \colhead{(u,g,r,i,z)} & \colhead{(Deg$^2$)} &  \colhead{Time (\%)} & \colhead{Surveys } 
 }
 \startdata 
Galactic & 21 & 3,300 & 0.6 & Galactic Spur Mini-Survey\\
Ecliptic Extragalactic & 62 & 3,100 & 0.8 & WFD extension, Ecliptic Survey\\
Space Optimal & 83 & 3,200 & 1.1 & 33\% WFD extension\\
Total & N/A & 9.600 & 2.5\\
\enddata
\end{deluxetable}

\vspace{.3in}

\begin{table}[ht]
    \centering
    \begin{tabular}{l|l|l|l}
        \toprule
        Properties & Importance \hspace{.3in} \\
        \midrule
        Image quality &     1\\
        Sky brightness &  3\\
        Individual image depth &  2 \\
        Co-added image depth &   1\\
        Number of exposures in a visit   &  3 \\
        Number of visits (in a night)  &  1 \\ 
        Total number of visits &  1 \\
        Time between visits (in a night) &  1 \\
        Time between visits (between nights)  & 3  \\
        Long-term gaps between visits & 2\\
        Other (please add other constraints as needed) & 3\\
        \bottomrule
    \end{tabular}
    \caption{{\bf Constraint Rankings:} Summary of the relative importance of various survey strategy constraints. Please rank the importance of each of these considerations, from 1=very important, 2=somewhat important, 3=not important. If a given constraint depends on other parameters in the table, but these other parameters are not important in themselves, please only mark the final constraint as important. For example, individual image depth depends on image quality, sky brightness, and number of exposures in a visit; if your science depends on the individual image depth but not directly on the other parameters, individual image depth would be `1' and the other parameters could be marked as `3', giving us the most flexibility when determining the composition of a visit, for example.}
        \label{tab:obs_constraints}
\end{table}

\pagebreak

\subsection{Technical trades}
{\it What is the effect of a trade-off between your requested survey footprint (area) and requested co-added depth or number of visits?}
\\

The transient and galactic stream cases scale with area.  The Cosmology and extragalactic science cases prefer the space optimal \emph{Euclid} area, followed by the low-galactic-extinction area in the ecliptic.  Science cases are reduced following Table \ref{tab:Nvisit} with most science cases thresholding at the proposed depths. \\
\\
    
{\it If not requesting a specific timing of visits, what is the effect of a trade-off between the uniformity of observations and the frequency of observations in time? e.g. a `rolling cadence' increases the frequency of visits during a short time period at the cost of fewer visits the rest of the time, making the overall sampling less uniform.}
\\

The cadence is very flexible with only lose constraints noted in the sections above. \\
\\
{\it What is the effect of a trade-off on the exposure time and number of visits (e.g. increasing the individual image depth but decreasing the overall number of visits)?}
\\

For the RR Lyrae and galactic streams proper motion cases the number of visits matters as stated.  For all other cases its irrelevant. \\
\\
{\it What is the effect of a trade-off between uniformity in number of visits and co-added depth? Is there any benefit to real-time exposure time optimization to obtain nearly constant single-visit limiting depth?}
\\

There is little benefit to matching single visit depths. 
\\
\\
{\it Are there any other potential trade-offs to consider when attempting to balance this proposal with others which may have similar but slightly different requests?}
\\

This survey is intended as a baseline with the tables highlighting the trades. 
\\
\\
\section{Performance Evaluation}

\begin{itemize}
\item Overall Tables \ref{tab:Nvisit} and \ref{tab:NvisitSurvey} give the threshold for loss of science cases at each depth/area. 

\item For the overall galactic stream cases the metric is the total sky coverage of LSST.  As detailed in the LSST science book, the proper motion precision scales primarily with sensitivity and the time baseline.  The two visit depths proposed reach required sensitivity to achieve the individual epoch precision, so the primary metric is the length of time between the first and last epoch.  

\item The cosmology case scales as the area covered by both LSST and a given survey (\emph{Euclid}/DESI).

\end{itemize}

\vspace{.6in}

\section{Special Data Processing}
None.

\section{References}
\bibliographystyle{hunsrt}
\begingroup
\renewcommand{\section}[2]{}%
\bibliography{ms}
\endgroup
\end{document}